\documentclass[conference]{IEEEtran}

\IEEEoverridecommandlockouts
\usepackage{cite}
\usepackage{amsmath,amssymb,amsfonts}
\usepackage{algorithmic}
\usepackage{float}
\usepackage{subfig}
\usepackage{graphicx}
\usepackage{xspace}
\usepackage{textcomp}
\usepackage{xcolor}
\def\BibTeX{{\rm B\kern-.05em{\sc i\kern-.025em b}\kern-.08em
    T\kern-.1667em\lower.7ex\hbox{E}\kern-.125emX}}

\begin{document}

\title{Robust End-to-End Speaker Verification Using EEG}
\author{
\IEEEauthorblockN{Yan Han}
\IEEEauthorblockA{\textit{Brain Machine Interface Lab} \\
\textit{The University of Texas at Austin}\\
Austin, Texas}
\and
\IEEEauthorblockN{Gautam Krishna}
\IEEEauthorblockA{\textit{Brain Machine Interface Lab} \\
\textit{The University of Texas at Austin}\\
Austin, Texas}
\and
\IEEEauthorblockN{Co Tran}
\IEEEauthorblockA{\textit{Brain Machine Interface Lab} \\
\textit{The University of Texas at Austin}\\
Austin, Texas}
\and
\IEEEauthorblockN{Mason Carnahan}
\IEEEauthorblockA{\textit{Brain Machine Interface Lab} \\
\textit{The University of Texas at Austin}\\
Austin, Texas}
\and
\IEEEauthorblockN{Ahmed H Tewfik}
\IEEEauthorblockA{\textit{Brain Machine Interface Lab} \\
\textit{The University of Texas at Austin}\\
Austin, Texas}
}
\maketitle

\maketitle
\begin{abstract}
In this paper we demonstrate that performance of a speaker verification system can be improved by concatenating electroencephalography (EEG) signal features with speech signal features or only using EEG signal features. We use state-of-the-art end-to-end deep learning model for performing speaker verification and we demonstrate our results for noisy speech. Our results indicate that EEG signals can improve the robustness of speaker verification systems, especially in noiser environment.
\end{abstract}

\section{Introduction}
Speaker verification is the process of verifying whether an utterance belongs to a specific speaker, based on that speaker’s known utterances. Speaker verification systems are used as authentication system in many voice activated technologies, for example in applications like voice match for Google home. In our work we focused on text independent speaker verification \cite{bimbot2004tutorial,kinnunen2010overview}. Deep learning based speaker verification systems \cite{heigold2016end,wan2018generalized,variani2014deep,chen2015locally,zhang2016end,sadjadi2016ibm} are getting popular this days and such systems have improved the performance of speaker verification systems.

Even though deep learning models have improved the state-of-the-art performance of speaker verification systems, their performance is degraded in presence of background noise and they are prone to attacks in the form of voice mimicking by the adversaries who are interested in breaching the authentication system.

We propose to use electroencephalography (EEG) signals to address these challenges. EEG is a non-invasive way of measuring electrical activity of human brain. In \cite{krishna2019speech} we demonstrated that EEG features can help automatic speech recognition (ASR) systems to overcome the performance loss in presence of background noise. Further, prior works explained in the references \cite{marcel2007person,paranjape2001electroencephalogram,palaniappan2003new,poulos1999person,riera2008unobtrusive,campisi2014brain,la2014human} show that EEG pattern in every individual is unique and it can be used for biometric identification. Motivated by these prior results, we used EEG features to improve the performance of speaker verification systems operating, compared with using acoustic features under two different levels of noisy conditions. The use of EEG features to improve the robustness of speaker verification system is also motivated by the unique robustness to environmental artifacts exhibited by the human auditory cortex \cite{yang1991auditory,mesgarani2011speech}. Speaker verification using EEG will also help people with speaking disabilities like broken speech to use voice authentication technologies with better accuracy, thereby improving technology accessibility.

For this work we used the end-to-end speaker verification model explained in \cite{wan2018generalized} as it is the current state-of-the-art model for speaker verification. Major contribution of this paper is the demonstration of improving robustness of speaker verification systems using EEG features.

The rest of this paper is structured as follows. Section \ref{2} introduces the speaker verification model used in our work, while Section \ref{3}, \ref{4} and \ref{5} describe details of the EEG database, feature extraction and preprocessing method for the data set. Experimental results and analysis are presented in Section \ref{6}, and finally, we conclude the paper in Section \ref{7}.

\section{Speaker Verification Model}
\label{2}
We used generalized end-to-end model introduced in \cite{wan2018generalized} for performing text independent speaker verification. We used Google's Tensorflow deep learning library for implementing the model.

The model consists of a single layer of Long short term memory (LSTM) \cite{hochreiter1997long} or Gated recurrent unit (GRU) \cite{chung2014empirical} cell with 128 hidden units followed by a dense layer which is a fully connected linear layer followed by a softmax activation. During training phase, an utterance $i$ of an enrollment candidate $X$ is passed to the LSTM or GRU cell and L2 normalization is applied to the dense layer output to derive the embedding or d vector for that utterance $i$. Similarly d vectors are derived for all the utterances of the enrolment candidate $X$. Since different utterances can have different lengths, we used dynamic recurrent neural network (RNN) cell of tensorflow. The speaker model or centroid for $X$ is defined as the average value of all the d vectors obtained for all the enrollment utterances of $X$.

Consider a set of enrollment candidates \{$X_1$,$X_2$, $\cdots \cdots$, $X_n$\} and let us assume each $X_i$ has $t$ number of utterances. Now as per our earlier definition we can build a set of centroids as \{$C_1$, $C_2$, $\cdots \cdots$, $C_n$\} where $C_i$ is the centroid for $X_i$. Now a two dimensional similarity matrix with $n\times t$ rows and $n$ columns is constructed, that computes the cosine similarity between all $C_i$'s and d vectors corresponding to utterances from all $X_i$'s. The cosine similarity contains learnable parameters that act as scaling factors as explained in \cite{wan2018generalized}. A softmax layer is applied on the similarity matrix. We used the generalized end-to-end softmax loss explained in \cite{wan2018generalized} as the loss function for the model. For any ($C_i$, $d_i$) pair, where $d_i$ is a d vector corresponding to an utterance from any $X_i$, the final softmax layer in the model outputs 1 if both $C_i$ and $d_i$ are derived from the same $X_i$, otherwise it outputs 0. Batch size was set to one and gradient descent optimizer Adam \cite{kingma2014adam} was used.

During test time, along with the enrollment utterances from the test set, evaluation utterances are also passed to the trained model. The similarity is calculated between the centroids derived from the enrollment utterances from test set and d vectors corresponding to evaluation utterances. Figure \ref{fig2} shows the training methodology and Figure \ref{fig3} shows the testing methodology of the verification model. Figure \ref{fig2} is adapted from \cite{wan2018generalized}. By the term utterance through out this paper, we refer to the Mel-frequency cepstral coefficients (MFCC 13) or EEG features or concatenation of MFCC 13 and EEG features, depending on how the speaker verification model was trained. 
\begin{figure}[h]
\begin{center}
\includegraphics[height=3cm,width=0.25\textwidth,trim={1cm 1cm 1cm 0.1cm},clip]{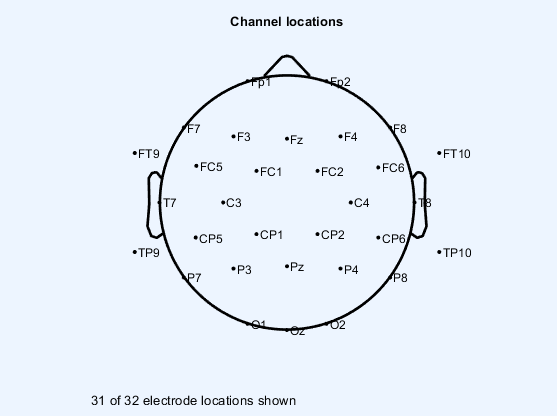}
\caption{EEG channel locations for the cap used in our experiments} 
\label{fig1}
\end{center}
\end{figure}

\section{Design of Experiments for building the database}
\label{3}

For this work, we have built two database of synchronized speech-EEG signals. For the data set A, 10 subjects from UT Austin (5 of them are female) took part in the experiment. While for the data set B, 8 subjects from UT Austin (3 of them are female) took part in the experiment. Since this work is about speaker verification, we don't require all the subjects to be native speakers. In our case, two of them are Chinese native speakers. And all subjects were graduate, undergraduate students in their early twenties.

For the both data sets, the subjects were asked to speaker the first 30 sentences from the USC-TIMIT database \cite{narayanan2014real} and their synchronized speech-EEG signals were recorded. The process was repeated for two more times. For the data set A, we had 30 speech-EEG recording samples for each sentence, and the level of background noise is 40 dB, generated by lab air conditioner fan. For the data set B, we had 24 speech-EEG recording samples for each sentence, and the level of background noise is 65 dB, generated by music of lab computer.

The Brain Vision EEG recording hardware is used in the experiment. The Figure \ref{fig1} shows our EEG cap, from the figure, we can find that it contains 32 wet EEG electrodes with one electrode as ground included. The EEG sensor topological location mapping is obtained from EEGLab\cite{delorme2004eeglab}, which is based on standard 10-20 EEG sensor placement method for 32 electrodes.

During the model training phase, for data set A, speech-EEG data of 8 subjects is used, with the remaining 2 subjects' data as test set. While for data set B, speech-EEG data of 6 subjects is used for training, with remaining 2 subjects' data as test set.






\section{EEG and Speech feature extraction details}
\label{4}

For preprocessing of EEG we followed the same method as described by the authors in \cite{yang1992auditory, krishna2019speech}. EEG signals were sampled at 1000Hz and a fourth order IIR band pass filter with cut off frequencies 0.1Hz and 70Hz was applied. A notch filter with cut off frequency 60 Hz was used to remove the power line noise. EEGlab’s \cite{delorme2004eeglab} Independent component analysis (ICA) toolbox was used to remove other biological signal artifacts like electrocardiography (ECG), electromyography (EMG), electrooculography (EOG) etc from the EEG signals. We extracted five features the same as the ones used by the authors in \cite{yang1992auditory, krishna2019speech}, namely root mean square, zero crossing rate, moving window average, kurtosis and power spectral entropy with frequency bands value equal to none (power per band was same as the power spectral density). All EEG features were extracted at a sampling frequency of 100 Hz. For this feature set, EEG feature dimension was 31(channels) $\times$ 5 or 155.

The recorded speech signal was sampled at 16KHz frequency. Mel-frequency cepstrum (MFCC) was extracted as features for speech signal. We extracted MFCC 13 features. In order to avoid seq2seq problem, the sampling frequency of MFCC features were also fixed at 100 Hz the same as it of EEG features. 

\begin{figure*}[h]
\begin{center}
\includegraphics[height=5cm,width=1\textwidth,trim={0.1cm 0.1cm 0.1cm 0.1cm},clip]{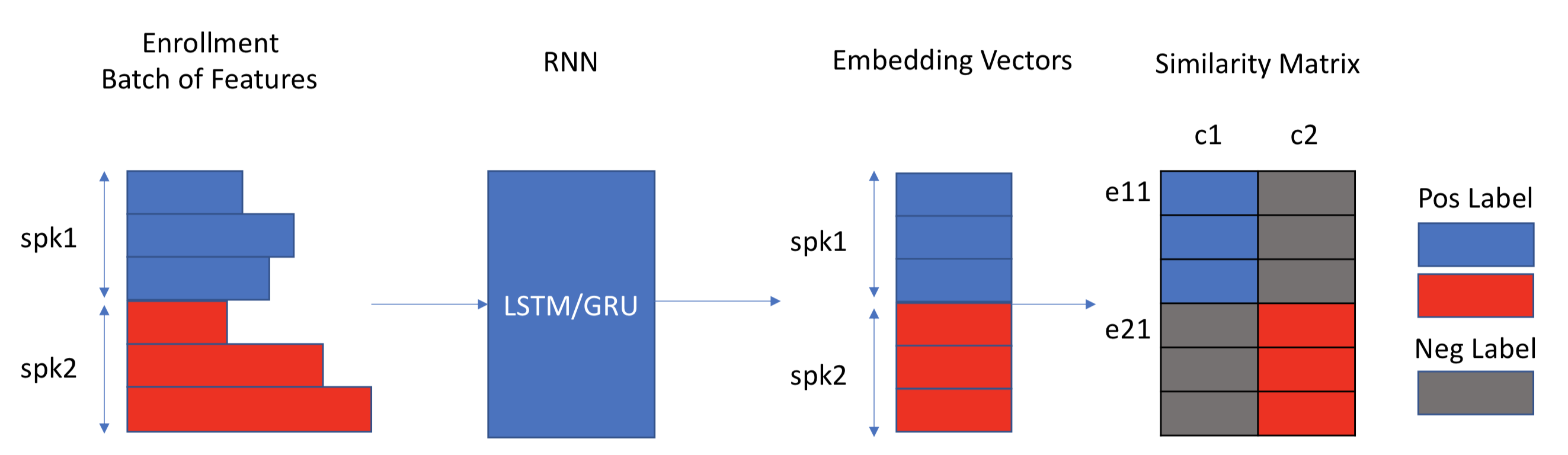}
\caption{Training method for Speaker Verification} 
\label{fig2}
\end{center}
\end{figure*}

\section{EEG Feature Dimension Reduction Algorithm Details}
\label{5}

We used non linear dimension reduction methods to denoise the EEG feature space. The tool we used for this purpose was Kernel Principle Component Analysis (KPCA)\cite{mika1999kernel}. We plotted cumulative explained variance versus number of components to identify the right feature dimension as shown in Figure \ref{fig4}. We used KPCA with polynomial kernel of degree 3\cite{yang1992auditory, krishna2019speech}. The cumulative explained variance plot is not supported by the library for KPCA as KPCA projects features into different feature space, hence for getting explained variance plot we used normal PCA but after identifying the right dimension we used KPCA to perform dimension reductions.

\begin{figure*}[h]
\begin{center}
\includegraphics[height=8cm,width=1\textwidth,trim={0.1cm 0.1cm 0.1cm 0.1cm},clip]{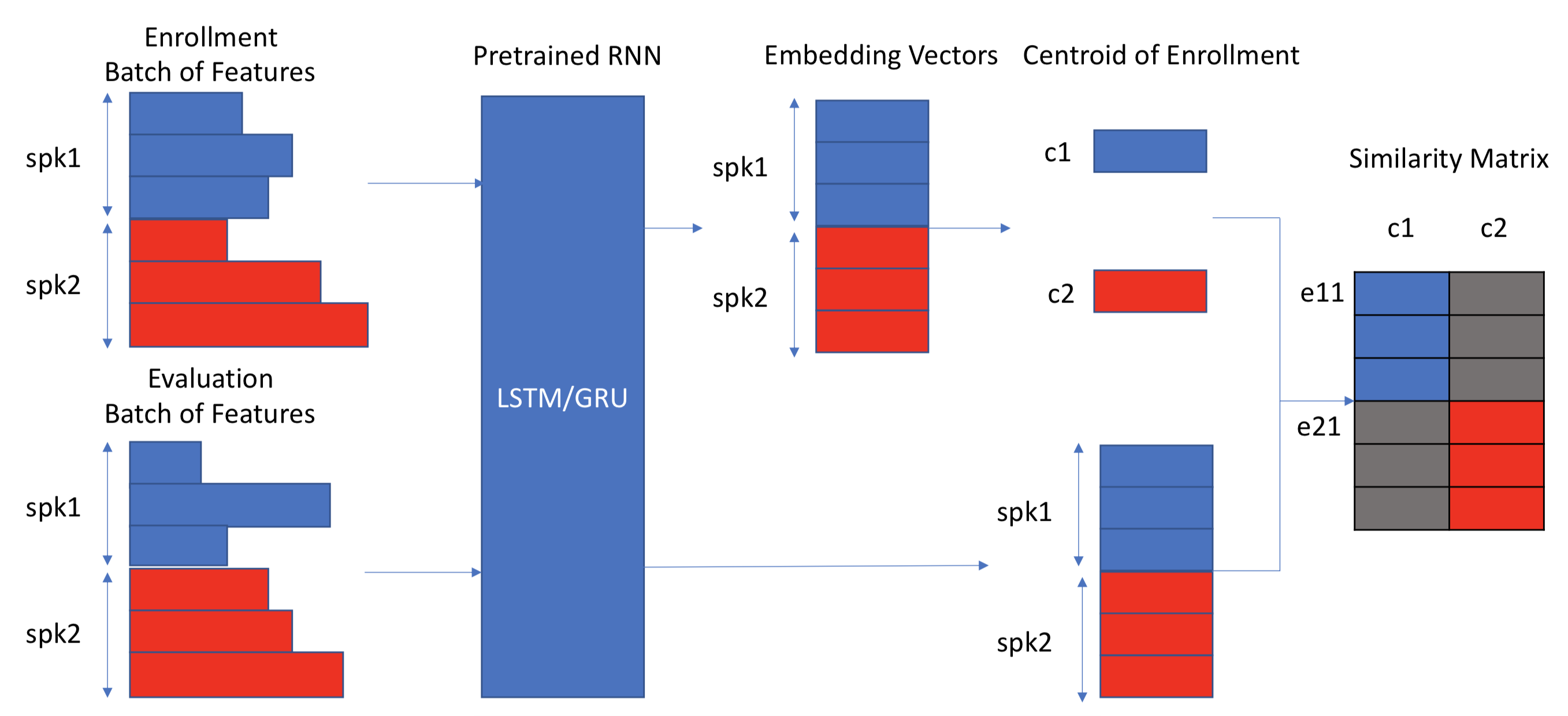}
\caption{Testing method for Speaker Verification} 
\label{fig3}
\end{center}
\end{figure*}
\begin{figure}[h]
\centering
\includegraphics[height=5cm, width=0.4
\textwidth,trim={0.1cm 0.1cm 0.1cm 0.1cm},clip]{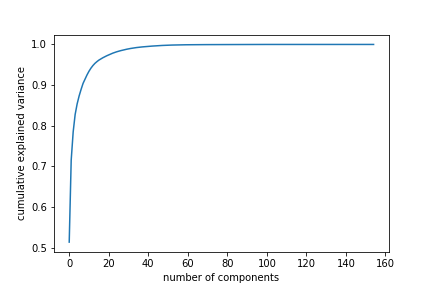}
\caption{Explained variance plot}
\label{fig4}
\end{figure}

\section{Results}
\label{6}
As explained earlier, for data set A we used data from first 8 subjects or 720 sentence utterances as training set, last two subjects or 180 sentence utterances as testing set. For data set B we used data from first 6 subjects or 540 sentence utterances as training set, last two subjects or 180 sentence utterances as testing set. There are 90 utterances per each subject. 
\begin{figure}[h]
\centering
\includegraphics[height=5cm, width=0.4
\textwidth,trim={0.1cm 0.1cm 0.1cm 0.1cm},clip]{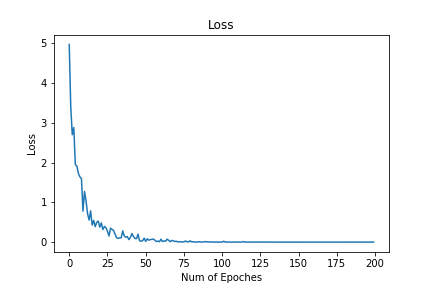}
\caption{Training Loss}
\label{fig5}
\end{figure}

Equal error rate (EER) defined in \cite{heigold2016end} was used as the evaluation metric to evaluate the model. Table \ref{table1} shows the obtained results on the less noisy data set A when we used LSTM as the RNN cell in the verification model and results clearly indicate that combining EEG features with acoustic features help in reducing the EER. In the table results are shown for number of sentence = \{3, 5, 7, 10, 15, 20, 30\}. Number of sentence = 3 implies during training, for first training step, features (MFCC or MFCC+EEG or EEG) corresponding to  first 3 sentence utterances are selected randomly from any two subjects from the training set and during second training step, features corresponding to next 3 sentence utterances are selected randomly from any two subjects and so on till the 90 sentence utterance is selected from the training set. So for this example, 30 training steps constitute one epoch. The model is trained for sufficient number of epochs to make sure it sees all the utterances present in the training set. In every epoch any two subjects are randomly chosen. 

During test time we have data from two subjects and number of sentences equal to 3 implies, for the first testing step features corresponding to first 3 sentence utterances from subject 1 and subject 2 are selected as enrollment utterances and features corresponding to next 3 sentence utterances from subject 1 and subject 2 are selected as evaluation utterances and EER is calculated. For second testing step, the features used as evaluation utterances in first step will become new enrollment utterances and features corresponding to next 3 sentence utterances from subjects 1, 2 will become new evaluation utterance and new EER is calculated. So for this example,there will be a total of 30 testing steps. EER value of 10\%, 7\% or 3\% for number of sentences equal to 3 as seen from the Table \ref{table1}, corresponds to the average of all the 30 testing step EER values. Similarly model was trained and tested for number of sentences = \{3, 5, 7, 10, 15, 20, 30\}. As 20 is not a factor of 90, for number of sentences equal to 20, there are five training steps in one epoch, five testing steps but the last step will contain only the features corresponding to last 10 sentence utterances. Figure \ref{fig5} shows the training loss convergence for the LSTM model for number of sentences equal to 3 on data set A when trained using EEG features.  

Table \ref{table2} shows the test time result using GRU model for
Data Set A, Table \ref{table3} shows the test time result using LSTM
model for Data Set B and Table \ref{table4} shows the test time result
using GRU model for Data Set B. The results shows that GRU model is almost better than LSTM model in all cases. The reason is that GRU is less complex than LSTM, and more suitable for a small size data set like our data set. 
\begin{table}[!ht]
\centering
\begin{tabular}{|l|l|l|l|}
\hline
\textbf{\begin{tabular}[c]{@{}l@{}}Number\\ of Sentences\end{tabular}} & \textbf{\begin{tabular}[c]{@{}l@{}}MFCC\\ (EER \%)\end{tabular}} & \multicolumn{1}{c|}{\textbf{\begin{tabular}[c]{@{}c@{}}MFCC+EEG\\ (EER \%)\end{tabular}}} &
\textbf{\begin{tabular}[c]{@{}l@{}}EEG\\ (EER \%)\end{tabular}}\\ \hline
3                                                                      & 10 & 7                                                              & {\color[HTML]{333333} \textbf{3}}                                                        \\ \hline
5                                                                     & 12 & 10                                                              & {\color[HTML]{333333} \textbf{5}}                                                        \\ \hline
7                                                                     & 14 & 11                                                               & {\color[HTML]{333333} \textbf{6}}                                                        \\ \hline
10                                                                     & 17 & 13                                                               & {\color[HTML]{333333} \textbf{8}}                                                        \\ \hline
20                                                                     & 19 & 14                                                              & \textbf{9}                                                                               \\ \hline
30                                                      & 22 & 16
                                                        & \textbf{11} \\ \hline
\end{tabular}
\caption{EER on test set for Data set \textbf{A} (40 dB) using \textbf{LSTM} based model}
\label{table1}
\end{table}

As can be seen from the tables, the EER of only using MFCC features is more prone to be worse under increasing level of environment noise compared with using MFCC+EEG features or only using EEG features, which indicates that EEG features are more robust for speaker verification.

\begin{table}[!ht]
\centering
\begin{tabular}{|l|l|l|l|}
\hline
\textbf{\begin{tabular}[c]{@{}l@{}}Number\\ of Sentences\end{tabular}} & \textbf{\begin{tabular}[c]{@{}l@{}}MFCC\\ (EER \%)\end{tabular}} & \textbf{\begin{tabular}[c]{@{}l@{}}MFCC+EEG\\ (EER \%)\end{tabular}} &
\textbf{\begin{tabular}[c]{@{}l@{}}EEG\\ (EER \%)\end{tabular}}\\ \hline
3                                                                      & 8 & 5                                                              & {\color[HTML]{333333} \textbf{3}}                                                        \\ \hline
5                                                                     & 11 & 8                                                              & {\color[HTML]{333333} \textbf{4}}                                                        \\ \hline
7                                                                     & 14 & 10                                                               & {\color[HTML]{333333} \textbf{6}}                                                        \\ \hline
10                                                                     & 15 & 11                                                               & {\color[HTML]{333333} \textbf{7}}                                                        \\ \hline
20                                                                     & 17 & 12                                                              & \textbf{8}                                                                               \\ \hline
30                                                      & 19 & 14
                                                        & \textbf{10} \\ \hline
\end{tabular}
\caption{EER on test set for Data set \textbf{A} (40 dB) using \textbf{GRU} based model}
\label{table2}
\end{table}
\begin{table}[!ht]
\centering
\begin{tabular}{|l|l|l|l|}
\hline
\textbf{\begin{tabular}[c]{@{}l@{}}Number\\ of Sentences\end{tabular}} & \textbf{\begin{tabular}[c]{@{}l@{}}MFCC\\ (EER \%)\end{tabular}} & \textbf{\begin{tabular}[c]{@{}l@{}}MFCC+EEG\\ (EER \%)\end{tabular}} &
\textbf{\begin{tabular}[c]{@{}l@{}}EEG\\ (EER \%)\end{tabular}}\\ \hline
3                                                                      & 12 & 10                                                              & {\color[HTML]{333333} \textbf{4}}                                                        \\ \hline
5                                                                     & 15 & 12                                                              & {\color[HTML]{333333} \textbf{6}}                                                        \\ \hline
7                                                                     & 16 & 13                                                               & {\color[HTML]{333333} \textbf{7}}                                                        \\ \hline
10                                                                     & 20 & 15                                                               & {\color[HTML]{333333} \textbf{8}}                                                        \\ \hline
20                                                                     & 23 & 18                                                              & \textbf{9}                                                                               \\ \hline
30                                                      & 26 & 19
                                                        & \textbf{12} \\ \hline
\end{tabular}
\caption{EER on test set for Data set \textbf{B} (65 dB) using \textbf{LSTM} based model}
\label{table3}
\end{table}

\begin{table}[!ht]
\centering
\begin{tabular}{|l|l|l|l|}
\hline
\textbf{\begin{tabular}[c]{@{}l@{}}Number\\ of Sentences\end{tabular}} & \textbf{\begin{tabular}[c]{@{}l@{}}MFCC\\ (EER \%)\end{tabular}} & \textbf{\begin{tabular}[c]{@{}l@{}}MFCC+EEG\\ (EER \%)\end{tabular}} &
\textbf{\begin{tabular}[c]{@{}l@{}}EEG\\ (EER \%)\end{tabular}}\\ \hline
3                                                                      & 11 & 9                                                              & {\color[HTML]{333333} \textbf{4}}                                                        \\ \hline
5                                                                     & 14 & 10                                                              & {\color[HTML]{333333} \textbf{5}}                                                        \\ \hline
7                                                                     & 15 & 11                                                               & {\color[HTML]{333333} \textbf{7}}                                                        \\ \hline
10                                                                     & 18 & 16                                                               & {\color[HTML]{333333} \textbf{7}}                                                        \\ \hline
20                                                                     & 21 & 15                                                              & \textbf{8}                                                                               \\ \hline
30                                                      & 24 & 17
                                                        & \textbf{11} \\ \hline
\end{tabular}
\caption{EER on test set for Data set \textbf{B} (65 dB) using \textbf{GRU} based model}
\label{table4}
\end{table}

\section{Conclusions}
\label{7}
In this paper we demonstrated the feasibility of using EEG signals to improve the robustness of speaker verification systems operating in very noisy environment. We demonstrated that for the two levels of noisy data, 40 dB and 65 dB, combination of MFCC and EEG, or only EEG features always resulted in lower EER.

Our overall results indicate that EEG features are less affected by background noise compared with MFCC features, which indicates that they are helpful in improving robustness of verification systems operating in presence of high background noise.

We further plan to publish our speech EEG data base used in this work to help advancement of the research in this area. For future work, we will work on building a much larger speech EEG data base and validate the results for larger data base. 

\section{Acknowledgements}
We would like to thank Kerry Loader from Dell, Austin, TX for donating us the GPU to train the models used in this work.\\

\bibliographystyle{IEEEtran}

\bibliography{IEEEabrv,mybib}


\end{document}